\documentclass[aps,prl,twocolumn,superscriptaddress,showpacs,draft]{revtex4}
\input{epsf}

\newcommand{\rem}[1]{}

\begin{document}

\title{Quantum ratchets in dissipative chaotic systems}
\author{Gabriel G. Carlo}
\affiliation{Center for Nonlinear and Complex Systems, Universit\`a degli 
Studi dell'Insubria and Istituto Nazionale per la Fisica della Materia, 
Unit\`a di Como, Via Valleggio 11, 22100 Como, Italy}
\author{Giuliano Benenti} 
\affiliation{Center for Nonlinear and Complex Systems, Universit\`a degli 
Studi dell'Insubria and Istituto Nazionale per la Fisica della Materia, 
Unit\`a di Como, Via Valleggio 11, 22100 Como, Italy}
\author{Giulio Casati} 
\affiliation{Center for Nonlinear and Complex Systems, Universit\`a degli 
Studi dell'Insubria and Istituto Nazionale per la Fisica della Materia, 
Unit\`a di Como, Via Valleggio 11, 22100 Como, Italy}
\author{Dima L. Shepelyansky}
\affiliation{Laboratoire de Physique Th\'eorique, 
UMR 5152 du CNRS, Universit\'e Paul Sabatier, 
31062 Touluse Cedex 4, France}
\date{July 27, 2004; Revised: January 19, 2005}


\begin{abstract} 
Using the method of quantum trajectories 
we study a quantum chaotic dissipative ratchet appearing for
particles in a pulsed asymmetric potential in the presence of
a dissipative environment. 
The system is characterized by directed transport emerging from
a quantum strange attractor.
This model exhibits, in the limit of small effective Planck constant,
a transition from quantum to classical behavior, in agreement with the 
correspondence principle. 
We also discuss parameter values suitable for implementation of
the quantum ratchet effect with cold atoms in optical lattices.
\end{abstract}
\maketitle
		
Cold atoms exposed to time-dependent standing waves of light 
provide an ideal test bed to explore the features of the 
quantum dynamics of nonlinear systems.
They allowed the experimental investigation of several important
physical phenomena, such as dynamical localization
\cite{Moore,Delande}, decoherence \cite{Ammann}, 
quantum resonances \cite{dArcy}, and chaos assisted 
tunneling \cite{Steck,Hensinger}.
In particular, they made possible to realize experimentally
the quantum kicked rotor \cite{Moore,Ammann,Delande,dArcy},
a paradigmatic model in the field of quantum chaos 
\cite{Casati}. 
Furthermore, cold atoms in optical lattices 
allowed to demonstrate  the ratchet phenomenon 
\cite{Mennerat,Schiavoni,Meacher}, that is characterized by
directed transport in the absence of any net force.

Directed transport \cite{Hanggi,Reimann}
is found in periodic systems due to a broken spatial-temporal 
symmetry \cite{Flach}, for instance in presence of lattice asymmetry, 
unbiased periodic driving and dissipation. 
This phenomenon, also called ratchet effect, is of potential 
relevance in technological applications such as
rectifiers, pumps, particle separation devices,
molecular switches, and transistors.
Moreover, it is of great interest for the understanding
of molecular motors in biology \cite{Julicher}. 
In previous works, attention has been focused on 
systems with external noise, on chaotic dynamical systems 
with dissipation and on purely Hamiltonian systems 
(with mixed or chaotic phase space)
\cite{Jung,Klafter,Schanz,Monteiro1,Monteiro2,Seba,Gong}.  
In particular, the issue of current reversal has 
attracted much interest \cite{Hanggi2,Mateos,Barbi,Dan}. 
It was found that current reversal can be originated by 
a bifurcation from a chaotic to a periodic regime,
that is, by the transition from a strange attractor to
a simple one \cite{Mateos}. 

The presence of strange attractors is a characteristic 
feature of classical dissipative chaotic systems \cite{Ott}. 
In quantum mechanics, it was found that the fractal 
structure of the classical strange attractor is smoothed
on the scale of Planck's cell \cite{Graham}. 
It is therefore interesting to investigate how this phenomenon 
affects quantum ratchets.
Till present the 
effects of dissipation have been mainly analyzed  for
classical ratchets while for quantum motion the
theoretical studies have been concentrated on the
Hamiltonian case \cite{Schanz,Monteiro1,Monteiro2}. 
In this Letter, we study a quantum ratchet system which is in contact 
with a dissipative environment and whose 
classical counterpart is chaotic. The investigation is carried out 
for parameter values where a strange attractor is present.  
Using the quantum trajectories approach, we are able to 
simulate numerically the evolution of a fractal quantum ratchet 
emerging from a quantum strange attractor.
Moreover, we show that the ratchet effect is robust in 
presence of noise.
Finally, we consider parameter values which should allow 
the observation of quantum chaotic dissipative ratchets 
in cold atoms experiments.

We study a particle moving in one dimension 
[$x\in(-\infty,+\infty)$] in a periodic kicked asymmetric 
potential:
\begin{equation}
V(x,\tau)=k\left[\cos(x)+\frac{a}{2}\cos(2x+\phi)\right]
\sum_{m=-\infty}^{+\infty}\delta(\tau-m T),
\end{equation}
where $T$ is the kicking period. The evolution of the 
system in one period is described by the map
\begin{equation}
\left\{
\begin{array}{l}
\overline{n}=\gamma n + 
k(\sin(x)+a\sin(2x+\phi)),
\\
\overline{x}=x+T\overline{n},
\end{array}
\right.
\label{dissmap} 
\end{equation}
where $n$ is the momentum variable conjugated to $x$
and $\gamma$ is the dissipation parameter, describing a
velocity proportional damping. We have $0\le \gamma \le 1$;
the limiting cases $\gamma=0$ and $\gamma=1$ correspond
to overdamping and Hamiltonian evolution, respectively. 
Introducing the rescaled momentum variable $p=Tn$, 
one can see that classical dynamics depends on the 
parameter $K=kT$ (not on $k$ and $T$ separately). 
We note that at $a=0$ the classical model reduces to the
Zaslavsky map \cite{Zaslavsky}. However, due to the symmetry
of the potential, the ratchet effect is absent in this case
(see discussion below).
The quantum model is obtained by means of the usual quantization
rules \cite{Casati}: 
$x\to \hat{x}$, $n\to \hat{n}=-i (d/dx)$ (we set $\hbar=1$).
Since $[\hat{x},\hat{p}]=iT$, the effective Planck constant 
is $\hbar_{\rm eff}=T$. The classical limit corresponds to 
$\hbar_{\rm eff}\to 0$, while keeping $K=\hbar_{\rm eff} k$ constant.  

In order to simulate a dissipative environment in the quantum 
model we consider a master equation in the Lindblad form 
\cite{Lindblad} for the density operator $\hat{\rho}$ of the 
system. We have
\begin{equation}
\dot{\hat{\rho}} = -i 
[\hat{H}_s,\hat{\rho}] - \frac{1}{2} \sum_{\mu=1}^2 
\{\hat{L}_{\mu}^{\dag} \hat{L}_{\mu},\hat{\rho}\}+
\sum_{\mu=1}^2 \hat{L}_{\mu} \hat{\rho} \hat{L}_{\mu}^{\dag},
\label{lindblad}
\end{equation}
where $\hat{H}_s=\hat{n}^2/2+V(\hat{x},\tau)$ is the system
Hamiltonian, 
$\hat{L}_{\mu}$ are the Lindblad operators, 
and \{\,,\,\} denotes the anticommutator. 
We assume that dissipation 
is described by the lowering operators
\begin{equation}
\begin{array}{l}
\hat{L}_1 = g \sum_n \sqrt{n+1} \; |n \rangle \, \langle n+1|,\\
\hat{L}_2 = g \sum_n \sqrt{n+1} \; |-n \rangle \, \langle -n-1|,
\end{array}
\end{equation} 
with $n=0,1,...$ \cite{grahamnote}. 
Requiring that at short times $\langle p \rangle $  
evolves like in the classical case, as it should be 
according to the Ehrenfest theorem, 
we obtain $g=\sqrt{-\ln \gamma}$.

The first two terms of Eq.~(\ref{lindblad}) can be regarded as the 
evolution governed by an effective non-Hermitian Hamiltonian, 
$\hat{H}_{\rm eff}=\hat{H}_s+i\hat{W}$, with 
$\hat{W}=-1/2 
\sum_{\mu}\hat{L}_{\mu}^{\dag}\hat{L}_{\mu}$. In turn, 
the last term is responsible for the so-called quantum jumps.
Taking an initial pure state $|\phi(\tau_0)\rangle$, 
the jump probabilities $dp_{\mu}$ in an infinitesimal time
$d\tau$ are defined by 
$
dp_{\mu}\!=\!\langle \phi(\tau_0)| \hat{L}_{\mu}^{\dag} 
\hat{L}_{\mu} |\phi(\tau_0)\rangle d\tau,
$
and the new states after the jumps by 
$|\phi_{\mu}\rangle = \hat{L}_{\mu} 
|\phi(\tau_0)\rangle/||\hat{L}_{\mu} |\phi(\tau_0)\rangle||$. 
With probability 
$dp_{\mu}$ a jump occurs and the system is left in the state 
$|\phi_{\mu}\rangle$. With probability $1-\sum_{\mu} dp_{\mu}$ there 
are no jumps and the system evolves according to the effective 
Hamiltonian $\hat{H}_{\rm eff}$.
In order to simulate the master equation (\ref{lindblad}), we have used 
the quantum trajectories approach \cite{DalibardCastin}. 
We start from a pure state $|\phi(\tau_0)\rangle$ 
and, at intervals $d\tau$, 
we perform the following evaluation. 
We choose a random number $\epsilon$ 
from a uniform distribution in the unit interval $[0,1]$. 
If $\epsilon < dp$, where $dp=\sum_{\mu=1}^2 dp_{\mu}$, 
the system jumps to one of the states $|\phi_{\mu}\rangle$ 
(to $|\phi_1\rangle$ if $0 \le \epsilon \le dp_1$ or to 
$|\phi_2\rangle$ if $dp_1 < \epsilon \le dp_1+dp_2$). 
On the other hand, if $\epsilon > dp$, the evolution with 
the non-Hermitian Hamiltonian $\hat{H}_{\rm eff}$ takes place, ending 
up in a state that we call $|\phi_0\rangle$. In 
both circumstances we renormalize the state. 
We repeat this process as many times as 
$n_{\rm steps}=(\Delta \tau)/d\tau$ where $\Delta \tau$ is the whole 
elapsed time during the evolution. 
Note that we must take $d\tau$ much smaller than the time scales relevant
for the evolution of the open quantum system under investigation.
In our simulations, $d\tau$ is inversely 
proportional to $\hbar_{\rm eff}$. Given any observable 
$A$, we can write the mean value $\langle A \rangle_\tau={\rm Tr}
[\hat{A} \hat{\rho}(\tau)]$ as the average over 
${\cal N}$ trajectories: 
\begin{equation}
\langle A \rangle_\tau = 
\lim_{{\cal N}\to \infty}
\frac{1}{{\cal N}} \sum_{i=1}^{{\cal N}} \langle \phi_i(\tau)| \hat{A} 
| \phi_i(\tau) \rangle.
\end{equation}
We found that a reasonable amount of trajectories 
(${\cal N} \simeq  100-500$) is sufficient in 
order to obtain a satisfactory statistical convergence. 
We used up to $3^{8}$ quantum levels to investigate
the transition from quantum to classical behavior.

\begin{figure}
\centerline{\epsfxsize=9cm\epsffile{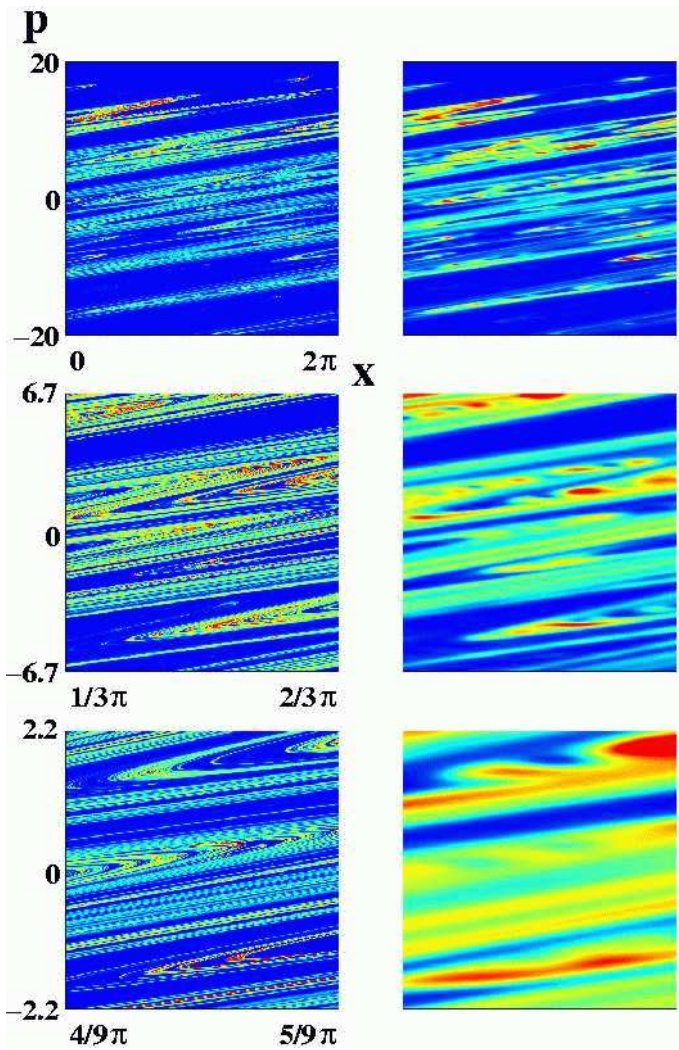}}
\caption{(Color) Phase space pictures for 
$K=7$, $\gamma=0.7$, $\phi=\pi/2$, $a=0.7$,
after $100$ kicks: classical Poincar\'e sections (left) 
and quantum Husimi functions at $\hbar_{\rm eff}=0.012$ (right).
In the upper row, the displayed region is given by 
$p \in [-20,20]$ and $x \in [0,2 \pi)$ (note that, to
draw the attractor, $x$ is taken modulus $2\pi$).
Magnifications of these plots are shown in the second and third 
rows (the area is reduced by a factor $1/9$ and 
$1/81$, respectively). 
The color is proportional to the density: blue for zero and 
red for maximal density.}
\label{fig1}
\end{figure}

The appearance of a strange attractor in our model is 
shown in the phase space portraits of 
Fig.~\ref{fig1}, for parameter values 
$K=7$, $\gamma=0.7$, $\phi=\pi/2$, $a=0.7$ \cite{attractornote}.
The left panel (classical mechanics) corresponds to a 
Poincare section constructed from $10^7$ random initial 
conditions uniformly distributed in the area
$x\in [0,2\pi)$, $p\in [-\pi;\pi]$, after $100$
iterations of map (\ref{dissmap}).   
From top to bottom we zoom in, showing the fractal 
structure of the attractor down to smaller and smaller
scales. 
On the right column (quantum mechanics, 
$\hbar_{\rm eff}=0.012$) we show the corresponding  
Husimi functions \cite{Husimi}
averaged over $N_i=5$ initial conditions 
(randomly selected $p$ values in the same interval as before),
considering ${\cal N}=160$ trajectories for each initial
condition. 
We can see a good agreement between the classical and the quantum 
phase space portraits. Though the quantum version is less detailed 
the main classical patterns are well reproduced. On the other hand, a 
magnification of a small part of the phase space
shows that the resolution of the quantum picture is 
limited by the uncertainty principle.

\begin{figure}
\centerline{\epsfxsize=8cm\epsffile{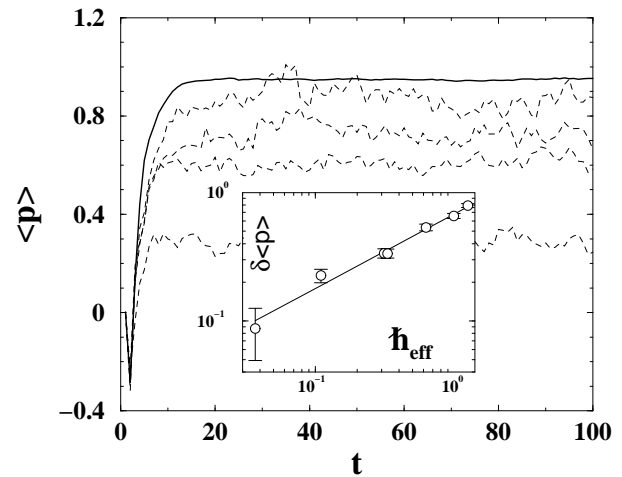}}
\caption{Average momentum $\langle p \rangle$ as a function
of time $t$ (measured in number of kicks), with same parameter 
values as in Fig.~\ref{fig1}.
The solid curve corresponds to the classical limit, 
while the other curves are quantum results at (from 
bottom to top) $\hbar_{\rm eff}=0.99,0.33,0.11,0.037$.
Each quantum curve is obtained from $N_i=60$ initial 
conditions and ${\cal N}=480$ trajectories for each 
initial conditions. Inset: scaling of 
$\delta\langle p \rangle$ with $\hbar_{\rm eff}$
(circles).
The power-law fit $\delta \langle p \rangle \propto 
\hbar_{\rm eff}^\nu$ gives $\nu\approx 0.6$ (full line).}
\label{fig2} 
\end{figure}

The attractor in Fig.~\ref{fig1} is strongly asymmetric, 
suggesting directed transport, that is, $\langle p \rangle \ne 0$.
This is confirmed by the numerical data of Fig.~\ref{fig2}, 
where $\langle p \rangle$ is shown as a function of time,
both in the classical and in the quantum case 
(for $\hbar_{\rm eff}=0.037,0.11,0.33,0.99$). 
In the classical case, $\langle p \rangle$ saturates after a 
very short time scale (approximately $10$ kicks) required for
the setting in of the strange attractor.
In the quantum case, a gradual approach to the classical 
limit can be clearly seen as $\hbar_{\rm eff}$ is
reduced, in agreement with the correspondence principle
The difference $\delta \langle p \rangle$ between the classical 
and quantum saturation currents is shown as a function of 
$\hbar_{\rm eff}$ in the inset of Fig.~\ref{fig2}. In the 
same inset, we also show the numerical fit 
$\delta \langle p \rangle\propto \hbar_{\rm eff}^\nu$, 
with the fitting constant $\nu\approx 0.6$.
We see that the quantum fluctuations lead to a smaller
value of the directed current as compared to the 
classical case.

\begin{figure}
\centerline{\epsfxsize=8cm\epsffile{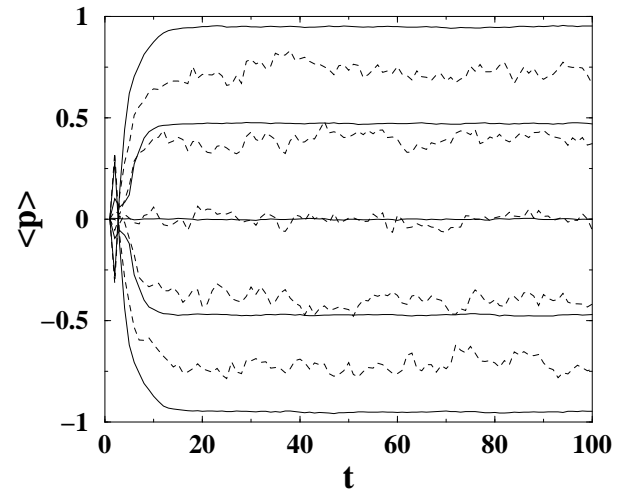}}
\caption{Average momentum versus time, 
for (from top to bottom) $\phi=\pi/2,2\pi/5,0,-2\pi/5,-\pi/2$,
in the classical (full curves) and in the quantum case
(dashed curves, $\hbar_{\rm eff}=0.11$). The other parameter
values are as in Figs.~\ref{fig1} and \ref{fig2}.} 
\label{fig3}
\end{figure}

In Fig.~\ref{fig3}, we show that it is possible to control
the direction of transport by varying the phase $\phi$. 
We observe that $\langle p \rangle_{-\phi} = 
- \langle p \rangle_\phi$. This is due to the fact that 
the potential $V(x,\tau)$ has the symmetry 
$V_\phi(x,\tau)=V_{-\phi}(-x,\tau)$. Therefore, the 
current can be reversed by changing $\phi\to -\phi$.
In particular, there is a space symmetry for $\phi=n \pi$ 
[in these cases, $V(x,\tau)=V(-x,\tau)]$, which implies
zero net current. This is confirmed, at $\phi=0$, from
the numerical data of Fig.~\ref{fig3}.  

We point out that quantum directed transport is observed in our 
model also for parameter values where the classical 
dynamics is characterized by a simple attractor (fixed points) 
instead of a strange one. Moreover, we have explored a different
dissipation mechanism ({\it i.e.}, loss of probability
for $|p|$ larger than some threshold value $p_{\rm th}$
after each kick \cite{Maspero}) and found that
this case is also characterized by directed transport and a 
quantum strange repeller \cite{notemaspero}. 
This is a further demonstration that 
quantum ratchets and asymmetric strange attractors and
repellers are
a typical feature of open quantum systems which are chaotic in 
the classical limit. 

\begin{figure}
\centerline{\epsfxsize=8cm\epsffile{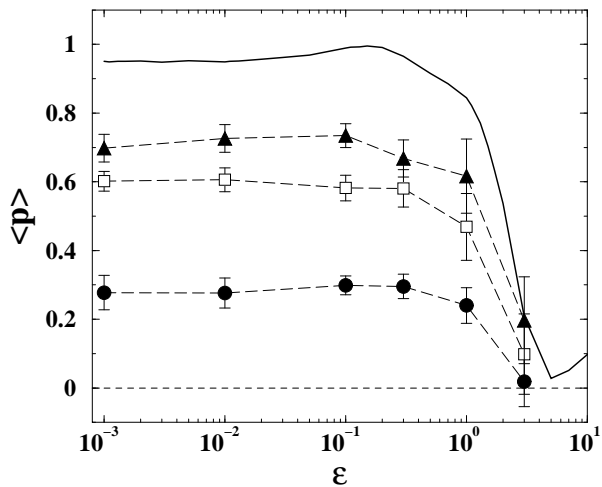}}
\caption{Average momentum versus noise strength, 
in the classical (full curve) and in the quantum case,
for $\hbar_{\rm eff}=0.99$ (circles), $0.33$ (squares),
and $0.11$ (triangles). The other parameter values are as
in Figs.~\ref{fig1},\ref{fig2}.} 
\label{fig4}
\end{figure}

It is important to analyze the stability of our ratchet 
phenomenon in respect to noise effects. 
For this purpose, we introduce memoryless fluctuations 
in the kicking strength: $K\to K_\epsilon(t)=K+\epsilon(t)$,
where the noise values $\epsilon(t)$ are uniformly and 
randomly distributed in the interval $[-\epsilon,+\epsilon]$.
The dependence of the classical and quantum currents
on $\epsilon$ is shown in Fig.~\ref{fig4}. 
We can see that directed transport survives, both in the 
classical and in the quantum case, up to a noise strength
$\epsilon$ of the order of the kicking strength $K$.
We stress that the robustness of our ratchet model is 
in contrasts with the behavior of the quantum Hamiltonian 
ratchets discussed in Refs.~\cite{Schanz,Monteiro1}. 
Indeed, noise eliminates the ratchet effect in 
Hamiltonian systems at large time scales, while in our 
case the ratchet is produced by dissipation and remains 
stable with respect to introduction of noise. 
We also note that the reduction of ratchet current with
noise (Fig.~\ref{fig4}) gives a qualitative reason
for reduction of the current with increase of
$\hbar_{eff}$ (Fig.~\ref{fig2}) which leads to
growth of quantum fluctuations.

The model discussed here can be realized 
experimentally with cold atoms in a periodic standing wave of light,
with parameters similar to those of 
Refs.~\cite{Moore,Delande,Ammann,dArcy,Steck}. 
For example, it is possible to use sodium atoms in a laser
field with wave length $\lambda_L=589$ nm \cite{Moore}, which creates 
effective pulsed potential with classical chaos parameter
$K=\sqrt{2\pi} \alpha \Omega_{\rm eff} \omega_r T^2$
and $\hbar_{\rm eff}=8 \omega_r T$. 
Here $T$ is the pulse periodicity, 
$\Omega_{\rm eff}$ the effective Rabi frequency,
$\omega_r=\hbar k_L^2/2M$ the recoil frequency  
($k_L=1/\lambda_L$, $M$ atomic mass) and $\alpha$
is the fraction of Gaussian pulse duration in units 
of pulse period. For typical parameter conditions of
Ref. \cite{Moore}, it is possible to take 
$\Omega_{\rm eff}/2\pi=75$ MHz, 
$T=0.8$ $\mu$s, $\alpha=0.05$, which gives
$\hbar_{\rm eff}\approx 1$, $K\approx 5$, being
similar to the case of Fig.2 \cite{fraction}. 
A dissipative friction force can be created by additional
laser fields which for sodium atoms can easily produce  
a dissipation rate $2\beta\approx 4\times 10^5$ s${}^{-1}$
\cite{Chu} giving $1-\gamma=2\beta T\approx 0.3$.
We also note that 
the Husimi function can in principle be measured from a state 
reconstruction technique \cite{Raizen}. This could allow 
the experimental observation of a quantum strange ratchet attractor.  

We acknowledge fruitful discussions with Marcos Saraceno. 
This work was supported in part by EU (IST-FET-EDIQIP) and 
NSA-ARDA (ARO contract No. DAAD19-02-1-0086).



\begin{thebibliography}{99}
\bibitem{grahamnote} 
These Lindblad operators can be obtained, similarly to 
\cite{Graham}, by considering the interaction between the 
system and a bosonic bath. The master equation (\ref{lindblad})
is then derived, at zero temperature, in the usual weak 
coupling and Markov approximations.
\bibitem{attractornote} 
There are also other 
similar parameter values where the system exhibits a simple 
or chaotic attractor.
\bibitem{notemaspero}
That is to say, a repelling chaotic limit set.
\bibitem{fraction}
At $\hbar_{\rm eff}\approx 1$ the ratchet current is 
$\langle p \rangle/\hbar_{\rm eff}\approx 0.3$ (see Fig.2), that 
corresponds to momentum equal to 0.3, expressed in units of double
photon recoil, $2 \hbar k_L$. This is significantly
larger than the experimental resolution equal to 0.03 \cite{Steck}.
\end{thebibliography}
\end{document}